\documentclass[twocolumn,twoside]{IEEEtran/IEEEtran}
\usepackage{times,helvet}
\usepackage{amsfonts}
\usepackage{amsmath}
\usepackage{amssymb}
\usepackage{wrapfig,booktabs}
\usepackage{hyperref}
\usepackage{enumitem}
\usepackage{graphicx}
\usepackage{rotating}
\usepackage{ulem}
\usepackage[T1]{fontenc}
\usepackage{tikz}
\usetikzlibrary{shapes,arrows}

\definecolor{olive}{rgb}{0.3, 0.8, .1}
\definecolor{lightblue}{rgb}{0, 0.5, 1}

\newcommand{\dd}{\mathsf{\,d}}

\newcommand{\sx}{\mathsf{x}}
\newcommand{\sy}{\mathsf{y}}
\newcommand{\sz}{\mathsf{z}}
\newcommand{\xp}{\textrm{E}}
\newcommand{\var}{\textrm{Var}}
\newcommand{\cov}{\textrm{Cov}}
\newcommand{\dg}{^\circ}
\newcommand{\fe}[1]{\hat{\sigma}_{#1}^2}

\newcommand{\subsubsubsection}[1]{

\vspace{2mm}

\noindent
-- \textit{#1}

}
\newcommand{\corad}[1]{#1}
\newcommand{\cordl}[1]{}

\newcommand{\engad}[1]{#1}
\newcommand{\engdl}[1]{}

\title{Three-Cornered Hat and Groslambert Covariance:\\
A first attempt to assess the uncertainty domains}

\author{Fran\c{c}ois Vernotte and \'Eric Lantz
\thanks{F. Vernotte is with UTINAM, UMR 6213 CNRS, Observatory THETA, Universit\'e Bourgogne Franche-Comt\'e, 41 bis avenue de l'Observatoire, 25010 Besan\c{c}on -- France}
\thanks{E. Lantz is with Institut FEMTO-ST, D\'epartement d'Optique P. M. Duffieux, UMR 6174 CNRS, Universit\'e Bourgogne Franche-Comt\'e, 15b Avenue des Montboucons, 25030 Besan\c{c}on -- France}
}

\begin{document}
\setcounter{page}{1}
\maketitle

\begin{abstract}
The three-cornered hat method and the Groslambert Covariance are very often used to estimate the frequency stability of each individual oscillator in a set of three oscillators by comparing them in pairs. However, no rigorous method to assess the uncertainties over their estimates has yet been formulated. In order to overcome this lack, this paper will first study the direct problem, i.e. the calculation of the statistics of the clock stability estimates by assuming known values of the true clock stabilities and then will propose a first attempt to solve the inverse problem, i.e. the assessment of a confidence interval over the true clock stabilities by assuming known values of the clock stability estimates. We show that this method is reliable from 5 Equivalent Degrees of Freedom (EDF) and beyond.
\end{abstract}

\section{Introduction}
The ``Three-Cornered Hat method'' was introduced by Gray and Allan in 1974 \cite{gray1974} to estimate the frequency stability of each individual oscillator in a set of three oscillators by comparing them in pairs. This method has been proved to be very efficient if the underlying assumption of the oscillator independence is fulfilled. Another approach, based on covariances, was proposed by Fest, Groslambert and Gagnepain in 1983 \cite{fest1983}. This method gives very similar results to such an extent that both methods were considered to be perfectly equivalent until very recently. However, we showed in 2016 that the latter, renamed the Groslambert Covariance (GCov), has the advantage to reject the noise measurement, i.e. the counter noise \cite{vernotte2016}. The use of GCov is becoming widespread and dedicated measuring instruments \corad{are} begin\corad{ning} to appear \cite{6ChPM, calosso2018}.

However, except \cordl{a few approximative} \corad{
approximated error bars 
 valid only for small integration times \cite{ekstrom2006}} or limited guidelines  \cordl{(e.g. \cite{vernotte2016}, {\S} III.E and III.F)} 
 \corad{\cite{vernotte2016}}, 
 no rigorous method to assess confidence intervals 
\cordl{or error bars}
 over the three-cornered hat or GCov estimates has yet been formulated. Nevertheless, it is an important issue, especially since negative variance estimates may be obtained by these methods, because these estimates are computed by calculating differences. Moreover, the uncertainty over each clock stability estimate strongly depends on the stabilities of all the clocks.

The aim of this paper is to study the statistics of the three-cornered hat/GCov method in order to find a way for computing confidence intervals over the true clock stabilities. After a quick reminder of both three-cornered hat and GCov methods, we will first address the so-called ``direct problem'', i.e. the calculation of the probability distribution of the clock stability estimates by knowing the true clock stabilities. Then, 
 we will propose a first attempt of solving the ``inverse problem'', i.e. assessing a confidence interval over the true clock stabilities by knowing the clock stability estimates obtained by the three-cornered hat/GCov method.
 
\section{Statement of the problem}
\subsection{Clock comparison}
Three-cornered hat as well as Groslambert Covariance rely both on simultaneous comparisons of 3 clocks $A$, $B$ and $C$ in pairs \cite{vernotte2016}. The clocks are assumed to be uncorrelated.

	\subsubsection{Time and frequency quantities}
In this paper we refer to time and frequency quantities such as phase time $\sx$, fractional frequency $\sy$, $\sz$ (see below)\ldots of oscillators. We indicate the signals from oscillators with capital letters ($A$, $B$, $C$\ldots) that are used as subscript\engad{s} of the related quantities, i.e. $\sy_A$ is \engad{the} fractional frequency of \engad{the} oscillator $A$. 
The measurement number $k$ can also be used as subscript.

	\subsubsection{Variances and covariances\label{sec:def_y}}
AVAR is generally defined as
$$
\sigma_y^2=\frac{1}{2}\xp{\left[\left(\bar{\sy}_{k+1}-\bar{\sy}_k\right)^2\right]}
$$
where $\xp{\left[\cdot\right]}$ is the mathematical expectation, and
$
\bar{\sy}_k=\frac{1}{\tau}\int_{t_k}^{t_{k+1}}\sy(t)\dd{t}
$
with $t_{k+1}=t_k+\tau$.

In order to simplify the notations, let us define the quantity $\sz_k=(\corad{\bar{\sy}}_{k+1}-\corad{\bar{\sy}}_k)/\sqrt{2}$. AVAR may then be written as
$
\sigma_y^2(\tau)=\xp{\left[\sz_k^2\right]}.
$
The associated estimator is then\footnote{In this paper, the symbol $\hat\cdot$ stands for the estimate of the quantity which is below.}
$$
\hat{\sigma}_y^2(\tau)=\frac{1}{M}\sum_{k=1}^M \sz_k^2\label{eq:est_AVAR}
$$
where $M$ \engad{is} the number of different $\sz_k$ in a data run of length $T$\corad{, whether AVAR is calculated with or without overlapping}. 


	\subsubsection{GCov vs 3-cornered hat}
The classical 3-cornered hat relies on the assumption of independence of the channel noises and of the oscillators $A$, $B$, $C$. The intercomparison of $A$ and $B$ is measured by $\sz_{AB}=\sz_B-\sz_A$ and its variance is
$
\sigma_{AB}^2=\sigma_A^2+\sigma_B^2
$
where all dependence\engad{s} on $\tau$ are omitted. Similarly, $\sigma_{BC}^2=\sigma_B^2+\sigma_C^2$ and $\sigma_{CA}^2=\sigma_C^2+\sigma_A^2$.
The three-cornered hat method uses the following property:
\corad{$
\sigma_{A}^2=\frac{1}{2}\left(\sigma_{AB}^2-\sigma_{BC}^2+\sigma_{CA}^2\right).
$}

On the other hand, the variance of oscillator $A$ may be estimated by using the covariance of the inter-comparison of $A$ and $B$ as well as \engad{of} the inter-comparison of $A$ and $C$. This leads to the Groslambert covariance
$\mathrm{GCov}_{A}=\xp\left[(\sz_A-\sz_B)(\sz_A-\sz_C)\right]$.

As mentioned above, these two approaches are almost equivalent\engad{,} but GCov is not polluted by the measurement noises since all cross-covariances are zero-mean  \cite{vernotte2016}. However, in this paper we will not distinguish these two approaches and will use one or the other method for mathematical derivations.

\subsection{Definition of the problem}
	\subsubsection{Measurements, model parameters and estimates\label{sec:choice_el_est}}
In the following, in order to simplify the notation, we will refer to $\sz_{PQ}$ and $\hat{\sigma}_{PQ}^2$ with $P, Q \in \left\{A, B, C \right \}$ to ensure the consistency of the notation regardless of the clock pair. The ``\textbf{elementary estimates}'' of the variance resulting from the comparison of the clocks $P$ and $Q$ is then
\begin{equation}
\hat{\sigma}_{PQ}^2=\frac{1}{M}\sum_{k=1}^M \sz_{PQ,k}^2.\label{eq:choice_el_est}
\end{equation}
Similarly, the \corad{Groslambert} covariance estimates will be denoted by 
\begin{equation}
\widehat{\mathrm{GCov}}_{PO,PQ}=\frac{1}{M}\sum_{k=1}^M \sz_{PO,k}\cdot \sz_{PQ,k}\label{eq:def_gcov}
\end{equation}
where $O, P, Q \in \left\{A, B, C \right \}$ and $M$ is the number of different $\sz_{PO,k}$ (or $\sz_{PQ,k}$) in a data run of length $T$. 

Finally, from these covariances or from the three-cornered hat, 
 we can compute the ``\textbf{final estimates}'' $\hat{\sigma}_{P}^2$ with $P \in \left\{A, B, C \right \}$.

	\subsubsection{Direct problem and inverse problem}
In order to assess the uncertainties over the estimation of the individual clock stabilities, we will have to distinguish two main issues:
\begin{itemize}
	\item The \textbf{direct problem} consists in calculating the statistics of the elementary estimates $\hat{\sigma}_{PQ}^2$, or final estimates $\hat{\sigma}_{P}^2$, knowing the model parameters $\sigma_{A}^2$, $\sigma_{B}^2$ and $\sigma_{C}^2$\engad{.} 
	\item The \textbf{inverse problem}, conversely, consists in calculating a confidence interval over each model parameter $\sigma_{P}^2$,  from the \cordl{measurements} \corad{final estimates} $\hat{\sigma}_{A}^2$, $\hat{\sigma}_{B}^2$ and $\hat{\sigma}_{C}^2$. Obviously, this last step is the true purpose of this paper.
\end{itemize}
This distinction corresponds to the main sections of this paper.

\section{Direct problem}
	\subsection{Calculation of two independent $\chi^2$ distributions}
		\subsubsection{Statistics of the $\hat{\sigma}_{PQ}^2$ estimates}
The two approaches described above are strictly equivalent when the counter noises are negligible\engad{.} \engdl{and t}\engad{T}his condition occurs for large integration time $\tau$. We will then alternatively use the 3-cornered hat or the GCov formalism for our demonstrations. Here we begin with \engad{the} 3-corner\corad{e}d hat but 
 the mathematical derivations below remain valid for the GCov method regardless of $\tau$. 

In the definition of the elementary estimates given by (\ref{eq:choice_el_est}), the number $M$ is a key element since it determines the number of Equivalent Degrees of F\corad{r}eedom (EDF) of these random variables (r.v.). If we use AVAR ``without overlapping'' on a White FM noise, the EDF is simply $\nu=T/\tau -1$ where $T$ is the total duration of the data-run. The use of Allan Variance with overlapping or its application to other noises than White FM will describe an identical statistics but with a different number of EDF. Since the number of EDFs is an independent parameter in this study, it will suffice to enter the number of EDF corresponding to the case treated in the obtained relationships, regardless \corad{of} the choice of the variance or the type of noise.

The $\sz_{PQ,k}$ measurements are Gaussian centered r.v. and Equation (\ref{eq:choice_el_est}) shows that $\hat{\sigma}_{PQ}^2$ is a r.v. that follows a $\chi^2$ distribution with $\nu$ degrees of freedom, which we will write $\chi_\nu^2$. The number of EDFs is here such that $1\leq\nu\leq M$ according to the correlation between the $\sz_{PQ,k}$  measurements. The precise determination of the number of EDFs does not fit into this study and is described in other publications (see for example \cite{greenhall}).

Therefore, the 3 estimates $\hat{\sigma}_{AB}^2$, $\hat{\sigma}_{BC}^2$ \cordl{et} \corad{and} $\hat{\sigma}_{CA}^2$ follow different $\chi^2$ distributions: $\hat{\sigma}_{AB}^2=k_{AB}\dot{\chi}_\nu^2$, $\hat{\sigma}_{BC}^2=k_{BC}\ddot{\chi}_\nu^2$ and $\hat{\sigma}_{CA}^2=k_{CA}\dddot{\chi}_\nu^2$ 
where the coefficients $k_{AB}, k_{BC}, k_{CA} \in \mathbb{R}^+$.
Since an oscillator is involved in two distributions, these distributions are correlated.

		\subsubsection{Vector formalization of the problem} 
To determine the statistics followed by the final estimates, we must  write them in the form of linear combinations of independent r.v.. Therefore, we have to determine an orthonormal basis of $N$ linearly independent vectors from $N'$ linearly dependent vectors, with $N<N'$. In our case, $N'=3$ (3 correlated $\chi^2$ \cordl{laws} \corad{r.v.}) and we will show in the following that $ N = 2 $ (2 independent $\chi^2$ \cordl{laws} \corad{r.v.}).

To simplify the problem, let us express (\ref{eq:choice_el_est}) in the case $M=1$ and thus $\nu = 1$ EDF:
$
\hat{\sigma}_{PQ}^2=\sz_{PQ}^2
$.
We can also consider that 
$\sz_{P} = (\corad{\bar{\sy}}_{P,2}-\corad{\bar{\sy}}_{P,1})/\sqrt{2}$ and therefore $\sz_{PQ} = \sz_{Q} -\sz_{P}$. Each $\sz_{P}$ follows a centered  \cordl{Gaussian} \corad{normal (Gaussian or Laplace-Gauss)} law $\textrm{LG}(0, \mathcal{Z}_P)$. Hence $\sz_{PQ} = \sz_{Q} -\sz_{P}$ follows a \engdl{a} centered  \engdl{Gaussian} \engad{normal} law of variance $\sigma_{PQ}^2 = \mathcal{Z}_P^2 + \mathcal{Z}_Q^2$.

Since the 3 clocks are independent, the quantities $\mathcal{Z}_P$  can be considered as the 3 coordinates of a vector in a 3-dimensional  space of basis $(\vec{e}_A, \vec{e}_B, \vec{e}_C)$. It is a vector space of \cordl{Laplace-Gauss} \corad{normal} laws since each of these normed basis vectors is characterized by a centered and reduced \cordl{Laplace-Gauss} \corad{normal} $\mathrm{LG}(0,1)$ independent in such a way that
$
\xp{\left[\vec{e}_P\cdot\vec{e}_Q\right]}=\delta_{P,Q}
$, where $\delta_{P,Q}$ represents the Kronecker symbol.

This vector space is thus endowed with a scalar product, denoted `$\cdot$', and a norm denoted `$||\cdot||^ 2$', defined by
$$
\left\{\begin{array}{l}
\vec{e}_P\cdot\vec{e}_Q=\dot{\mathrm{LG}}(0,1)\cdot\ddot{\mathrm{LG}}(0,1) \quad \textrm{if } P\neq Q\\
\vec{e}_P\cdot\vec{e}_P=||\vec{e}_P||^2=\chi_1^2
\end{array}\right.
$$
where $\chi_1^2$ is a r.v. following a $ \chi^2$ law with 1 degree of freedom and $\dot{\mathrm {LG}}(0,1)$ and $\ddot{\mathrm{LG}}(0,1) $ represent 2 independent \cordl{Laplace-Gauss} \corad{normal} \cordl{laws} \corad{r.v}. Their product therefore follows a Bessel distribution of mathematical expectation 0 and variance 1 \cite{sorin1968}, which is a special case of the variance-gamma distribution \corad{defined in} \cordl{that we will note V$\Gamma(1,0,1/2)$ (see} Appendix \ref{sec:annexe}\cordl{)}, \corad{with $\mathcal{A}=\mathcal{B}$ and $\nu=1$, and then $\theta=0$, $\eta=\lambda=\kappa=1/2$ in (\ref{eq:pdf})}.

Thus, the quantities $\sz_A$, $\sz_B$ \cordl{et} \corad{and} $\sz_C$ become vectors that we will write $\vec{A}$, $\vec{B}$ \cordl{et} \corad{and} $\vec{C}$ and which are defined by:
$\vec{A}=\mathcal{Z}_A \vec{e}_A = \left(\mathcal{Z}_A, 0, 0\right)^\textrm{T}$, 
$\vec{B}=\mathcal{Z}_B \vec{e}_B = \left(0, \mathcal{Z}_B, 0\right)^\textrm{T}$,
and $\vec{C}=\mathcal{Z}_C \vec{e}_C = \left(0, 0, \mathcal{Z}_C\right)^\textrm{T}$.

We can now express the variances and covariances of the primary oscillators through norms and scalar products of vectors: 
$
||\vec{B}||^2=
\mathcal{Z}_B^2 ||\vec{e}_B||^2=\mathcal{Z}_B^2\chi_1^2$. 
We see in particular that $\sigma_B^2=\xp[||\vec{B}||^2]=\mathcal{Z}_B^2$. As for the scalar product between two different basis vectors, its expectation is null and equal to the covariance between two independent oscillators. However, it is described by a variance-gamma distribution:
$
\vec{A}\cdot \vec{B}=
\mathcal{Z}_A\mathcal{Z}_B \vec{e}_A\cdot \vec{e}_B=\mathcal{Z}_A\mathcal{Z}_B \mathrm{V}\Gamma(1,0,1/2)
$
.

The clock comparisons $\sz_{AB}$, $\sz_{BC}$ \cordl{et} \corad{and} $\sz_{CA}$ also become vectors formed according to the following model:
$
\sz_{AB}=\sz_B-\sz_A=\overrightarrow{AB}=\vec{B}-\vec{A}=-\mathcal{Z}_A \vec{e}_A+\mathcal{Z}_B \vec{e}_B
$.

Similarly, the variances $\hat{\sigma}_{AB}^2$, $\hat{\sigma}_{BC}^2$ and $\hat{\sigma}_{CA}^2$ can be rewritten with these notations:
\begin{eqnarray}
\hat{\sigma}_{AB}^2&=&||\overrightarrow{AB}||^2=\mathcal{Z}_A^2||\vec{e}_A||^2+\mathcal{Z}_B^2||\vec{e}_B||^2\nonumber\\
&&-2\mathcal{Z}_A\mathcal{Z}_B\vec{e}_A\cdot \vec{e}_B\nonumber\\
&=&\mathcal{Z}_A^2\dot{\chi}^2_1+\mathcal{Z}_B^2\ddot{\chi}^2_1-2\mathcal{Z}_A\mathcal{Z}_B\mathrm{V}\Gamma(1,0,1/2)\nonumber
\end{eqnarray}

It is now easy to see that the 3 vectors $\overrightarrow{AB}$, $\overrightarrow{BC}$ \cordl{et} \corad{and} $\overrightarrow{CA}$ are linearly dependent since 
$\overrightarrow{AB}+\overrightarrow{BC}+\overrightarrow{CA}=\vec{0}$.

  Therefore, they all belong to a 2-dimensional subspace of the above defined 3-dimensional space. It is thus necessary to look for an orthonormal basis $(\vec{u}_1,\vec{u}_2)$ in which we can rewrite these 3 vectors.

		\subsubsection{Search for an independent basis} 
The Gram-Schmidt algorithm can be used to solve this problem.
Let us choose $\vec{u}_2 =\gamma \overrightarrow{CA}$ in such a way that $\xp\left[||\vec{u}_2||^2\right]=1$:
$$
||\vec{u}_2||^2 =
\gamma^2 \mathcal{Z}_A^2 \dot{\chi}_1^2+\gamma^2 \mathcal{Z}_C^2 \dddot{\chi}_1^2-2\gamma^2 \mathcal{Z}_A \mathcal{Z}_C \mathrm{V}\Gamma(1,0,1/2)
$$
and then
$
\xp\left[||\vec{u}_2||^2\right]=\gamma^2 \mathcal{Z}_A^2+\gamma^2 \mathcal{Z}_C^2=1
$
\engdl{whence} \engad{where}
\begin{equation}
\gamma=\frac{1}{\sqrt{\mathcal{Z}_A^2+\mathcal{Z}_C^2}}.\label{eq:def_gamma}
\end{equation}

We now need to find a second vector, $\vec{u}_1$, which is a linear combination of $\overrightarrow{AB}$ and $\overrightarrow{BC}$, perpendicular to $\vec{u}_2$\corad{, with a norm of unity mathematical expectation:} \cordl{and whose mathematical expectation of norm is 1:}
$$\left\{\begin{array}{l}
\vec{u}_1 = \alpha \overrightarrow{AB} - \beta \overrightarrow{BC}\\
\xp\left[\vec{u}_1\cdot \vec{u}_2\right]=-\alpha\gamma \mathcal{Z}_A^2+\beta\gamma \mathcal{Z}_C^2=0\\
\xp\left[||\vec{u}_1||^2\right]=\alpha^2\mathcal{Z}_A^2+(\alpha-\beta)^2\mathcal{Z}_B^2+\beta^2\mathcal{Z}_C^2=1.\\
\end{array}\right.
$$
The solution is:
\begin{equation}
\left\{\begin{array}{l}
\alpha=\frac{\mathcal{Z}_C^2}{\sqrt{\mathcal{Z}_A^4(\mathcal{Z}_B^2+\mathcal{Z}_C^2)+\mathcal{Z}_C^4(\mathcal{Z}_A^2+\mathcal{Z}_B^2)+2\mathcal{Z}_A^2\mathcal{Z}_B^2\mathcal{Z}_C^2}}\\
\beta=\frac{\mathcal{Z}_A^2}{\sqrt{\mathcal{Z}_A^4(\mathcal{Z}_B^2+\mathcal{Z}_C^2)+\mathcal{Z}_C^4(\mathcal{Z}_A^2+\mathcal{Z}_B^2)+2\mathcal{Z}_A^2\mathcal{Z}_B^2\mathcal{Z}_C^2}}\\
\end{array}\right.\label{eq:def_alpha_beta}
\end{equation}

We can now write the three vectors $\overrightarrow{AB}$, $\overrightarrow{BC}$ and $\overrightarrow{CA}$ as linear combinations of the basis vectors $(\vec{u}_1,\vec{u}_2)$:
\begin{equation}
\overrightarrow{CA}=\frac{1}{\gamma}\vec{u}_2.\label{eq:CA}
\end{equation}
Using the property $\overrightarrow{AB}+\overrightarrow{BC}+\overrightarrow{CA}=0$, we obtain:
\begin{eqnarray}
\vec{u}_1&=&\alpha \overrightarrow{AB} - \beta\overrightarrow{BC}=\alpha \overrightarrow{AB} - \beta\left(-\overrightarrow{AB}-\overrightarrow{CA}\right)\nonumber\\
&=&(\alpha+\beta)\overrightarrow{AB}+\frac{\beta}{\gamma}\vec{u}_2\nonumber
\end{eqnarray}
\engdl{whence it finally comes} \engad{leading finally to}:
\begin{equation}
\overrightarrow{AB}=\frac{1}{\alpha+\beta}\vec{u}_1-\frac{\beta}{\gamma(\alpha+\beta)}\vec{u}_2.\label{eq:AB}
\end{equation}
Similarly, we find:
\begin{equation}
\overrightarrow{BC}=-\frac{1}{\alpha+\beta}\vec{u}_1-\frac{\alpha}{\gamma(\alpha+\beta)}\vec{u}_2.\label{eq:BC}
\end{equation}

Thanks to the equations (\ref{eq:CA}), (\ref{eq:AB}) \cordl{et} \corad{and} (\ref{eq:BC}), we can estimate $\sigma_{B}^2$, for example\engdl{, thanks to} \engad{by using} $\widehat{\textrm{GCov}}_{BA,BC}$ :
\begin{eqnarray}
\widehat{\textrm{GCov}}_{BA,BC}&=& \overrightarrow{BA}\cdot \overrightarrow{BC}\nonumber\\
&=&\frac{1}{(\alpha+\beta)^2}\dot{\chi}_1^2-\frac{\alpha\beta}{\gamma^2(\alpha+\beta)^2}\ddot{\chi}_1^2\nonumber\\
&&+\frac{\alpha-\beta}{\gamma(\alpha+\beta)^2}\mathrm{V}\Gamma(1,0,1/2)\nonumber
\end{eqnarray}
where now the 2 $\chi^2$ \cordl{laws} \corad{r.v.} are totally independent. We note, however, that there is a third term of null mathematical expectation, but \cordl{whose} \corad{of which} variance is non-zero except in the particular case where $\alpha=\beta$, i.e. $ \mathcal{Z}_A = \mathcal{Z}_C$. In this case, we will have access to the distribution of $\hat\sigma_B^2$ since we know that the difference of two $\chi^2$ \cordl{laws} \corad{r.v.} corresponds to a r.v. \cordl{whose} \corad{with a} probability density \cordl{is} given by the equation (\ref{eq:pdf}) of Appendix \ref{sec:annexe}.

In the general case\engdl{, that is} $\alpha \neq \beta$, a Bessel distribution is added to the difference between the $\chi^2$ \cordl{laws} \corad{r.v.}, preventing us from calculating the density of $\hat\sigma_B^2$. This problem can be solved by rotating the basis $(\vec{u}_1, \vec{u}_2)$ to get a new basis $(\vec{V}_1, \vec{V}_2)$ so that the multiplicative factor of the scalar product $\vec{V}_1 \cdot \vec{V}_2$ is identically zero.


		\subsection{Rotation of the basis vectors\label{sec:rot_pb}}
It is thus necessary to find the eigenvector basis $(\vec{V}_1, \vec{V}_2)$ \engdl{so} \engad{such} that
\begin{eqnarray}
\hat\sigma_B^2&=&\frac{1}{(\alpha+\beta)^2}||\vec{u}_1||^2-\frac{\alpha\beta}{\gamma^2(\alpha+\beta)^2}||\vec{u}_2||^2\nonumber\\
&&+\frac{\alpha-\beta}{\gamma(\alpha+\beta)^2}\vec{u}_1\cdot \vec{u}_2\label{eq:avant_rot}\\
&=&\mathcal{A}||\vec{V}_1||^2-\mathcal{B}||\vec{V}_2||^2+\mathcal{C}\vec{V}_1\cdot \vec{V}_2\label{eq:apres_rot}\nonumber
\end{eqnarray}
with $\mathcal{C}=0$.
Let us simplify the notation of Equation (\ref{eq:avant_rot}): 
$
\hat\sigma_B^2=a||\vec{u}_1||^2-b||\vec{u}_2||^2+c\vec{u}_1\cdot \vec{u}_2
$
with
\begin{equation}
\left\{\begin{array}{l}
a = \displaystyle \frac{1}{(\alpha+\beta)^2}\\
b = \displaystyle \frac{\alpha\beta}{\gamma^2(\alpha+\beta)^2}\\
c = \displaystyle \frac{\alpha-\beta}{\gamma(\alpha+\beta)^2}.
\end{array}\right.\label{eq:def_abc}
\end{equation}
It is a quadratic form which associates a scalar $w_0$ to any vector $\vec{w}=w_1\vec{u}_1+w_2\vec{u}_2$ according to
\begin{equation}
w_0=\vec{w}^TQ\vec{w}\label{eq:quad_form}
\end{equation}
with 
$$
Q=\left(
\begin{array}{cc}
a & c/2\\
c/2 & -b
\end{array}\right).
$$
Diagonalizing the matrix $Q$ gives the eigenvalues:
\begin{equation}
\left\{\begin{array}{l}
\mathcal{A}=\frac{a-b+\sqrt{(a+b)^2+c^2}}{2}\vphantom{\frac{\frac{1}{2}}{\frac{1}{2}}}\\
\mathcal{B}=\frac{a-b\corad{-}\sqrt{(a+b)^2+c^2}}{2}\vphantom{\frac{\frac{1}{2}}{\frac{1}{2}}}.
\end{array}\right.\label{eq:defAB}
\end{equation}
and eigenvectors 
$
\vec{E}_A=\left(c/2, \mathcal{A}-a\right)^\textrm{T}$ and $\vec{E}_B=\left(\mathcal{B}+b, 
c/2\right)^\textrm{T}$.
The eigenvector matrix is $E=\left(\vec{E}_A\ \vec{E}_B\right)$
and the quadratic form of (\ref{eq:quad_form}) may be rewritten as
$
w_0=\vec{w}^TE\Lambda E^T\vec{w}
$.
Thus, this new formulation can be interpreted as a basis change in which a vector $\vec{w}$ \engdl{will be} \engad{is} transformed into a vector $\vec{W}$ according to:
\begin{equation}
\vec{W}=E^T\vec{w}.\label{eq:rot_w}
\end{equation}

The basis change of relation (\ref{eq:rot_w}) is a rotation of angle $\varphi$ 
 since
$$
E^T
=\left(\begin{array}{cc}
\cos \varphi & \sin \varphi\\
-\sin \varphi & \cos \varphi
\end{array}\right)\nonumber
$$
with
\begin{equation}
\varphi=\arctan\left[\left(-a-b+\sqrt{(a+b)^2+c^2}\right)/c\right]\label{eq:def_ABCD_phi}.
\end{equation}

This angle, although not essential for the calculation of basis vectors, proves to be a very useful indicator to analyze the data (see Section \ref{sec:mod_rot}).


	\subsection{Validation of the theoretical probability laws by Monte Carlo simulations}
		\subsubsection{Validation principle}
According to the preceding section, the probability density of $\hat\sigma_B^2$, equal to the difference of two independent  $\chi^2$ \cordl{laws} \corad{r.v.}, can now be calculated using the function $p(x)$ of the equation (\ref{eq:pdf}) of Appendix \ref{sec:annexe} by assigning the following values to the parameters of this function:
\begin{equation}
\left\{\begin{array}{l}
\corad{\eta=
\frac{\mathcal{A}+\mathcal{B}}{4\mathcal{A}\mathcal{B}}\vphantom{\frac{\frac{1}{2}}{\frac{1}{2}}}}\\
\corad{\theta=
\frac{\mathcal{A}-\mathcal{B}}{4\mathcal{A}\mathcal{B}}\vphantom{\frac{\frac{1}{2}}{\frac{1}{2}}}}\\
\kappa=
\sqrt{\eta^2-\theta^2}=\frac{1}{2\sqrt{\mathcal{A}\mathcal{B}}}\\
\lambda=
\nu/2\\
\mu=0
\end{array}\right.\label{eq:param_pdf}
\end{equation}
where $\mathcal{A}$ and $\mathcal{B}$ are the values given in \corad{(\ref{eq:defAB})} and $\nu$ the number of EDF of the considered $\chi^2$ laws. Indeed, if the above study was done for $\nu=1$ for reasons of simplification of the formalism, it remains perfectly valid regardless of the number of EDFs.

To verify the results of this model, we compared them with those given by a simulation that seems realistic: we chose to return to the frequency deviations of the individual clocks by simulating the quantities $\bar{\sy}_{A,k}$, $\bar{\sy}_{B,k}$ \cordl{et} \corad{and} $\bar{\sy}_{C,k}$ (see {\S} \ref{sec:def_y}). They were simulated by a centered \cordl{Laplace-Gauss} \corad{normal} \cordl{law} \corad{r.v.} (\textit{randn} function of \textit{Octave}, the \textit{Matlab} clone). It might be objected that we \engdl{are} \engad{were} simulating white noise while frequency deviations are \engad{much more} likely to be \engdl{much more} red noise samples. But what matters in this study is the Gaussian character of the probability law more than the power law of its spectral density. The only effect of the latter \engdl{will be to reduce} \engad{is reducing} the number of EDFs of the $\chi^2$ laws. 

			\subsubsubsection{Simulation algorithm}
The simulation algorithm follows these 6 steps
\renewcommand{\labelenumi}{S\arabic{enumi}:}
\begin{enumerate}
	\item Assignment of the 3 noise levels $\mathcal{Z}_P$ 
	\item Draw\engad{ing} of $3\times \nu$ samples $\bar{\sy}_{P,k}$ 
	\item Computation of the $3\times \nu$ differences $\bar{\sy}_{PQ,k}=\bar{\sy}_{Q,k}-\bar{\sy}_{P,k}$ 
	\item Computation of the 3 estimates $\hat\sigma_P^2=\widehat{\textrm{GCov}}_{PO,PQ}$ 
	\item Repetition $N=10^7$ times of \engad{the} steps S2 to S4 of this sequence.
	\item Draw\engad{ing} of the 3 histograms of the $\hat\sigma_P^2$.
\end{enumerate}
In all simulations, we chose a number of EDF $\nu=5$.

			\subsubsubsection{Modeling algorithm}
The modeling algorithm follows these 6 steps for each estimate $\hat\sigma_P^2$ 
\begin{enumerate}
	\item Assignment of the noise levels of the 3 clocks $\mathcal{Z}_A$, $\mathcal{Z}_B$ and $\mathcal{Z}_C$
	\item \textbf{Independent basis} 
	\begin{itemize}
		\item Computation of the coefficients $\alpha$, $\beta$ and $\gamma$ according to the relationships (\ref{eq:def_gamma}) and (\ref{eq:def_alpha_beta})
		\item Computation of the coefficients $a$, $b$ and $c$ according to (\ref{eq:def_abc})
	\end{itemize}
	\item \textbf{Vector rotation} (option) 
	\begin{itemize}
		\item Computation of the rotation angle $\varphi$ \engdl{thanks to} \engad{with} (\ref{eq:def_ABCD_phi})
		\item Computation of the coefficients $\mathcal{A}$, $\mathcal{B}$ and $\mathcal{C}$ \engdl{thanks to} \engad{with} (\ref{eq:defAB})
	\end{itemize}
	\item Computation of the coefficients $\eta$, $\theta$, $\kappa$ and $\lambda$ \engdl{thanks to} \engad{with} (\ref{eq:param_pdf})
	\item Plotting the probability density with (\ref{eq:pdf}).
	\item Repeat \engad{the} steps S2 to S5 for the other 2 clocks.
\end{enumerate} 

		\subsubsection{Modeling without the basis rotation\label{sec:stab_comp}}
All the histograms shown in the following figures have been normalized by the total number of draws and the width of the bins to get the average probability density over each bin.
 
\begin{figure}
\includegraphics[width=9cm]{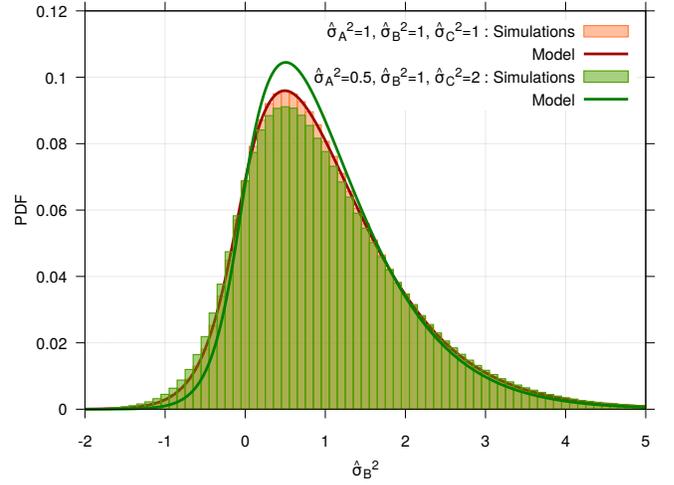}

\vspace{-6mm}

\caption{Probability densities of the final estimate $\hat{\sigma}_B^2$ obtained by the model without rotation (solid lines) and by simulations (boxes) for comparable stability clocks: $\sigma_A^2=\sigma_B^2=\sigma_C^2=1$ (red) ; B. $\sigma_A^2=0,5$, $\sigma_B^2=1$, $\sigma_C^2=2$ (green). The chosen number of EDFs was $\nu=5$.\label{fig:stab_comp}}
\end{figure}

As expected, when the noise levels of the 3 clocks are the same, the match between the model and the simulations is excellent  (figure \ref{fig:stab_comp} in red). 

On the other hand, when the differences in noise level are significant ($\sigma_C^2=2\sigma_B^2=4\sigma_A^2$), the gap becomes obvious, especially with regard to the measurement $\hat{\sigma}_B^2$ (see Figure \ref{fig:stab_comp} in green). It is indeed for this one that the disparity between the 2 other levels is the most important ($\sigma_C^2=8\sigma_A^2$) and the approximation $\mathcal{A}-\mathcal{B}\ll \mathcal{A}+\mathcal{B}$, necessary for the model without rotation to be valid, is no longer verified.

%

		\subsubsection{Modeling with the rotation\label{sec:mod_rot}}
Now, let's put the model with the rotation of the basis vectors on the test bench by comparing it to the simulations.
			\subsubsubsection{Clocks with comparable stabilities \cordl{and beyond}}

\begin{figure}[b]
\includegraphics[width=9cm]{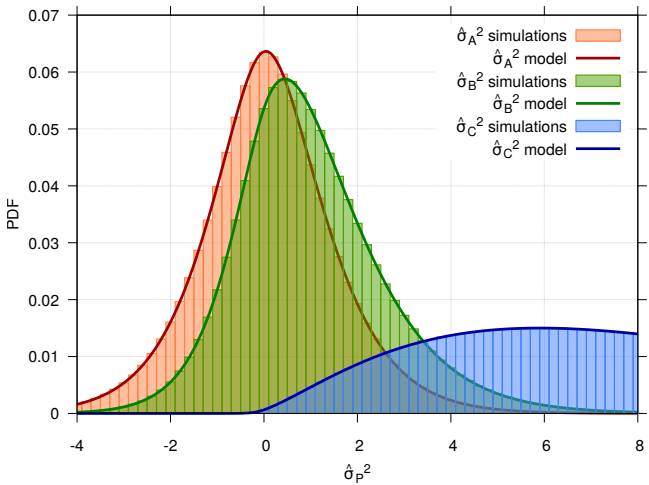}

\vspace{-6mm}

\caption{Probability densities of the final estimate $\hat{\sigma}_A^2$ (red), $\hat{\sigma}_B^2$ (green), $\hat{\sigma}_C^2$ (blue), obtained by the model with rotation (solid lines) and by the simulations (boxes). The stability of the clock ensemble was given by the following parameters:  $\sigma_A^2=0.1$, $\sigma_B^2=1$, $\sigma_C^2=10$.
 The chosen number of EDFs was $\nu=5$.\label{fig:stab_comp_rot}}
\end{figure}

At first, we took again the levels of stability tested with the model without rotation (see Section \ref{sec:stab_comp}) and we found in each case an excellent agreement between model and simulation. 

\begin{table}
\caption{Comparison of 3 clocks of comparable stabilities \cordl{and beyond} using the rotation of the eigenvectors. The chosen number of EDFs was $\nu=5$.\label{tab:stab_comp_rot}} 
\centering 
\begin{tabular}{c|cc|cc|cc}
\hline
 & \multicolumn{2}{|c|}{$\sigma_A^2=0.1$} & \multicolumn{2}{|c|}{$\sigma_B^2=1$} & \multicolumn{2}{|c}{$\sigma_C^2=10$}\\
\hline
Angle & \multicolumn{2}{|c|}{$27.43\dg$} & \multicolumn{2}{|c|}{$-34.93\dg$} & \multicolumn{2}{|c}{$7.49\dg$}\\
\hline
Fractile & Model & Simul. & Model & Simul. & Model & Simul. \\
2.5 \% & -2.894 & -2.893 & -1.773 & -1.775 & 1.428 & 1.428 \\
97.5 \% & 3.190 & 3.193 & 4.715 & 4.715 & 26.09 & 26.08\\
\hline
$P(\hat{\sigma}<0)$ & 47.5 \% & 47.4 \% & 26.6 \% & 26.6 \% & 0.06 \% & 0.06 \% \\
\hline
\end{tabular}
\end{table}


We can treat with the same success the case where there is a factor 100 between $\sigma_A^2$ and $\sigma_C^2$ (see Table \ref{tab:stab_comp_rot} and Figure \ref{fig:stab_comp_rot}). To be able to compare the results given by the model and the simulations, we give them with 4 significant digits in Table \ref{tab:stab_comp_rot}. Whether it be \engad{for} the fractiles \engad{(lines 4 and 5)} or \engad{for} the probability of getting a negative measurement \engad{(line 6)}, we can notice that the results agree at least up to the significant 3$^\textrm{\footnotesize rd}$ digit.

We obtain measurements that are almost symmetrically distributed around 0 for $\hat{\sigma}_A^2$, meaning that these measurements are \cordl{overwhelmed in} \corad{masked by} the measurement fluctuations  of the other clocks. Therefore, the stability of this clock is not measurable using the two clocks $B$ and $C$ which are much less stable. On the other hand, the stability of the least stable clock, $\sigma_C^2$, is relatively well determined (let us remind that with only 5 EDFs, the confidence intervals remain nevertheless very wide) since the rate of negative measurements is only 6 out of $10\,000 $. Finally, the determination of the stability $\sigma_B^2$ is intermediate between those of $\sigma_A^ 2$ and of $\sigma_C^2$.

			\subsubsubsection{One clock is much less stable or much more stable than the other\engad{s} \engdl{ones}}
We then examined the case where two of the clocks are of identical stabilities and the other is either 100 times more stable or 100 times less stable. Here also we found a full agreement between model and simulation.

To summarize, the model with the rotation of the basis vectors fits perfectly whatever the noise levels of the clocks \engdl{are}. Therefore, we can consider that we have solved the direct problem.

\section{Inverse problem}
\subsection{Principle of the method}
	\subsubsection{Inverse problem and parameter uncertainties}
The metrologist has to solve the inverse problem, i.e. the determination of a confidence interval for the true variances $\sigma_A^2$, $\sigma_B^2$ and $\sigma_C^2$, given a set of measurements and a priori information, i.e. any information known before the measurements. In this experimental world, the true variances appear as random variables, \cordl{whose} \corad{of which} the a posteriori \engad{probability} density \engdl{of probability}, i.e. probability density that takes into account the measurements, is determined by using the Bayes theorem: 
\begin{equation}
\left\{\begin{array}{l}
\displaystyle p\left(\sigma_A^2,\sigma_B^2,\sigma_C^2|\hat{\sigma}_A^2,\hat{\sigma}_B^2,\hat{\sigma}_C^2\right) \\
\\
\displaystyle \propto \pi\left(\sigma_A^2,\sigma_B^2,\sigma_C^2\right) p\left(\hat{\sigma}_A^2,\hat{\sigma}_B^2,\hat{\sigma}_C^2|\sigma_A^2,\sigma_B^2,\sigma_C^2\right)\\
\\
\displaystyle\iiint_0^\infty p\left(\sigma_A^2,\sigma_B^2,\sigma_C^2|\hat{\sigma}_A^2,\hat{\sigma}_B^2,\hat{\sigma}_C^2\right) \mathrm{d}\sigma_A^2 \mathrm{d}\sigma_B^2 \mathrm{d}\sigma_C^2= 1\label{eq:posterior}
\end{array}\right.
\end{equation}
where  $\pi(\theta)$ is the a priori probability density, named prior,  \engdl{on} \engad{of} a value $\theta$ \cite{bernardo87}, here the variances. Even if we have no a priori information on the variances, $\pi(\sigma_A^2,\sigma_B^2,\sigma_C^2)$ can be defined: a variance is a positive scale parameter\cite{lindley1958, lantz2013} and a prior reflecting no a priori knowledge (total ignorance)\label{sec:ignorance} is proportional to $\frac{1}{\theta}$, meaning that all orders of magnitudes have the same a priori probability. Because the oscillators are assumed to be independent, the prior of a triplet of variances is simply given by the product of the individual priors:
\begin{equation}
\pi(\sigma_A^2,\sigma_B^2,\sigma_C^2)\propto \frac{1}{\sigma_A^2 \sigma_B^2\sigma_C^2} \nonumber
\end{equation}

To compute the a posteriori probability of a variance triplet given by Eq. (\ref{eq:posterior}), it remains to calculate the probability $p(\hat{\sigma}_A^2,\hat{\sigma}_B^2,\hat{\sigma}_C^2|\sigma_A^2,\sigma_B^2,\sigma_C^2)$. This is a direct problem, that has been solved in the first part of this paper for a single estimate, for example $p(\hat{\sigma}_A^2|\sigma_A^2,\sigma_B^2,\sigma_C^2)$. Unfortunately, the three final estimates $\hat{\sigma}_A^2$, $\hat{\sigma}_B^2$, $\hat{\sigma}_C^2$ are not independent and the probability of a  triplet of estimate is not given by the product of the probabilities determined in the preceding section. This issue cannot be solved by using the elementary estimates defined in section (\ref{sec:choice_el_est}): $\hat{\sigma}_{AB}^2$, $\hat{\sigma}_{AC}^2$, $\hat{\sigma}_{BC}^2$, 
since they are neither independent (each oscillator participates in two of them). Before making probability products, we must find linear combinations of our three estimates that are independent, or, at least, uncorrelated. It is well known that these combinations are obtained  by applying the Karhunen-Loève (K.L.) transform to our original estimates, either elementary or final \cite{vernotte1990}. 
In the following lines, the principle of the K.L transform is recalled and applied to our specific case. Since we are supposed to know at this step the true variances, it is possible to calculate the true covariance matrix of our estimates (either $\hat{\sigma}_A^2,\hat{\sigma}_B^2,\hat{\sigma}_C^2$ or $\hat{\sigma}_{AB}^2$, $\hat{\sigma}_{AC}^2$, $\hat{\sigma}_{BC}^2$), 
and to find the rotation that \cordl{rends} \corad{renders} diagonal this covariance matrix. The calculation of this true covariance matrix is detail\cordl{l}ed in Section \ref{sec:covmat}.
The coefficients of the eigenvectors of this diagonalisation process are used  as weighting numbers to compute a new triplet of uncorrelated estimates. Actually, these K.L estimates do not obey \cordl{a} Gaussian statistics, since the original estimates follow the not trivial probability density law exposed in the preceding sections. However, we will approximate in the following this distribution by a Gaussian one. If we assume that this approximation is correct, the three K.L. estimates obey each \cordl{a} Gaussian statistics, with their three variances given by the diagonalized covariance matrix. Moreover, in the frame of this approximation, the K.L. estimates are independent and the probability density of a K.L. triplet is simply given by the product of the three probabilit\corad{y}\cordl{ies} densities. Because of the one to one correspondence between the K.L. triplet and the original estimates, this probability density is also proportional to the probability density of the triplet of original estimates. We expect that this Gaussian approximation becomes more accurate for a large number of measurements (more EDF). We will see in the following that this is correct, but the approximation can  be used even for a small number of measurements, at least 5 (see Section \ref{sec:acc_EDF}), at the price of a reasonable inaccuracy in the limits of the confidence intervals. 
This inaccuracy will be assessed in Section %
 \ref{sec:acc_inv_pb}. 

\subsubsection{Algorithm}\label{sec:algo}
We have to calculate the triple integral of Eq. (\ref{eq:posterior}) on several order of magnitudes. A direct calculation would lead to prohibitive computation times. We have  preferred to use a Monte-Carlo scheme with random sampling. This sampling ensures the observance of the total ignorance a priori law: the samples are chosen at random on a logarithmic scale \corad{in such a way that all orders of magnitude have the same probability (see the concept of total ignorance in \S \ref{sec:ignorance})}, independently for each variance. With a computation on 8 decades (between $10^{-5}$ and 1$0^3$) with $10^4$ sampling steps, $10^7$ samples proved to be sufficient to \cordl{rend} \corad{render} negligible the sampling error, in comparison with other inaccuracies. With the same sampling step, a direct calculation would involve $10^{12}$ cells.

 We work in the experimenter point of view: we assume that a triplet of estimates (either final or elementary) has been calculated from the $3\cdot m$ elementary measurements. These three numbers have three definite values that will be used in the calculations detailed below. The different steps of the calculation can be summarized as

\begin{enumerate}
	\item Choose at random a triplet of true variances, with a uniform probability on a logarithmic scale for each variance and independence between the three variances.
	\item Calculate for this triplet the covariance matrix of the estimate triplet, either using the final or the elementary estimates.
	\item Calculate the eigenvectors and eigenvalues of this covariance matrix
	\item Multiply the vector of the estimates by the matrix of these eigenvectors (K.L. transform).
	\item Perform the same operation for the vector of true variances.
	\item Calculate the probability density of each K.L estimate given the triplet of K.L. true variances: for each of the three K.L. variables, we assume a Gaussian \corad{normal} law of mean the K.L true variance and of variance the corresponding eigenvalue of the covariance matrix.\label{it:approxLG}
	\item Perform the product of these three probability densities. This only number will be associated in the following to the triplet of variances chosen at the first step of the algorithm.
	\item Repeat $10^7$ times the entire process.
	\item For each of the three variables, normalize the probability densities by dividing by their sum (sum of  $10^7$ values).
	\item Also for each of the three variables, sort the true variance values and calculate the cumulative density function by a partial sum on the associated normalized probability densities. 
	\item Determinate a confidence interval at $95 \%$ on each true variance from the corresponding cumulative density functions.
	\item Verify that the low limit of the confidence interval is meaningful\cordl{l}. For a Gaussian distribution, $99.7 \%$ of data are included in a confidence interval at $\pm 3 \sigma$. If the low limit of this $\pm 3 \sigma$ confidence interval (in logarithmic scale) is smaller than the low limit of the a priori range (here $10^{-5}$ ), we suspect (and have verified)  that the low limit of the smaller confidence interval calculated in the preceding step will depend on the low limit of the a priori range. If it occurs, we replace the low limit of the confidence interval by 0.
\end{enumerate}

 We have verified that employing the elementary or the final estimates gives exactly the same results. Moreover, the uncertainties due to the random character of the Monte-Carlo integration are less than $1 \%$ in relative value. Hence, the only non negligible cause of error in the algorithm is the Gaussian approximation. This error is assessed in the following paragraph (see Section \ref{sec:acc_inv_pb}). But before, it is useful to find the expected properties of the final estimates.
 
	\subsubsection{Properties of the final estimates\label{sec:finest_prop}} 
		\subsubsubsection{General properties}
\renewcommand{\labelenumi}{P\arabic{enumi}:}
\begin{enumerate}
	\item \textit{Only one final estimate may be negative.}\\
Demonstration: All elementary estimates are positive 
$
\fe{AB}=\fe{A}+\fe{B}>0 \ \textrm{and} \ \fe{BC}=\fe{B}+\fe{C}>0 \ \textrm{and} \ \fe{CA}=\fe{C}+\fe{A}>0.
$
Therefore, if two final estimates would be negative, at least one of the elementary estimate would be negative and this is impossible.
	\item \textit{If a final estimate is negative, its absolute value is smaller than the absolute values of the other two final estimates of the triplet.}\\
Demonstration: if $\fe{A}<0$, 
$\fe{AB}=\fe{A}+\fe{B}=-|\fe{A}|+|\fe{B}|>0$ 
$\Rightarrow$ $|\fe{B}|>|\fe{A}|$. Similarly, 
$\fe{CA}=\fe{C}+\fe{A}=|\fe{C}|-|\fe{A}|>0$
$\Rightarrow$ $|\fe{C}|>|\fe{A}|$.
\end{enumerate}
		\subsubsubsection{Case of $\nu=1$ EDF}
\begin{enumerate}
\setcounter{enumi}{2}
	\item 
$\displaystyle\fe{P}=-\frac{\fe{O}\fe{Q}}{\fe{O}+\fe{Q}}$ \textit{with 1 EDF
and $\left\{O, P, Q\right\}$ any circular permutation of $\left\{A, B, C \right\}$.}\\
Demonstration: 
With 1 EDF, Equation (\ref{eq:def_gcov}) becomes: $
\fe{P}=-\sz_{OP}\cdot \sz_{PQ}=-(\sz_P-\sz_O)(\sz_Q-\sz_P)$. Similarly, with 1 EDF Equation (\ref{eq:choice_el_est}) becomes :
$
\fe{PQ}=\sz_{PQ}^2=(\sz_Q-\sz_P)^2
$
Moreover 
$\fe{PQ}=\fe{P}+\fe{Q}$. Therefore
$$
-\frac{\fe{O}\fe{Q}}{\fe{O}+\fe{Q}}=-\frac{\fe{O}\fe{Q}}{\fe{OQ}}
$$
$$
=-\frac{(\sz_O-\sz_Q)(\sz_O-\sz_P)(\sz_Q-\sz_P)(\sz_Q-\sz_O)}{(\sz_Q-\sz_O)^2}
$$
$$
=-(\sz_P-\sz_O)(\sz_Q-\sz_P)=\fe{P}.
$$
\end{enumerate}

This latter propert\corad{y}\cordl{ies} is very important since the case of $\nu=1$ EDF occurs for the largest integration time, i.e. $\tau=T/2$ for AVAR and a dataset of duration $T$. For this largest $\tau$, Property 3 implies two major consequences:
\begin{itemize}
	\item one of the final estimates of the triplet and only one is necessar\corad{il}y negative 
	\item the knowledge of two final estimates of the triplet is enough since the third one can be deduced from them by using Property 3 (e.g. if $\fe{B}=\fe{C}=1$ then $\fe{A}=-1/2$).
\end{itemize}

\subsection{Accuracy of the inverse algorithm\label{sec:acc_inv_pb}}
	\subsubsection{Principle of the simulations}
In order to assess the accuracy of the method we propose, we compared it to Monte-Carlo simulations. The principle consists in randomly drawing parameter triplets $(\sigma_A^2, \sigma_B^2, \sigma_C^2)$, computing the corresponding final estimate triplet $(\hat{\sigma}_A^2, \hat{\sigma}_B^2, \hat{\sigma}_C^2)$ and keeping only the parameter triplets which yield  a previously given final estimate triplet $(\hat{\sigma}_A^2, \hat{\sigma}_B^2, \hat{\sigma}_C^2)=(A_0,B_0,C_0)$. Obviously, the random generation is of importance: as previously, we choose each element of each triplet independently according to an uniform probability law on a logarithmic scale. The pseudo-random numbers may vary within a large interval depending on the ($A_0,B_0,C_0)$ triplet and of the EDF, typically between $B_l=10^{-3}$ and $B_h=10^{+2}$ for $(1,1,1)$. Each simulation run stops when 10,000 achievements have been obtained.

	\subsubsection{Preliminary observations on the simulations}
\begin{figure}
\includegraphics[width=9cm]{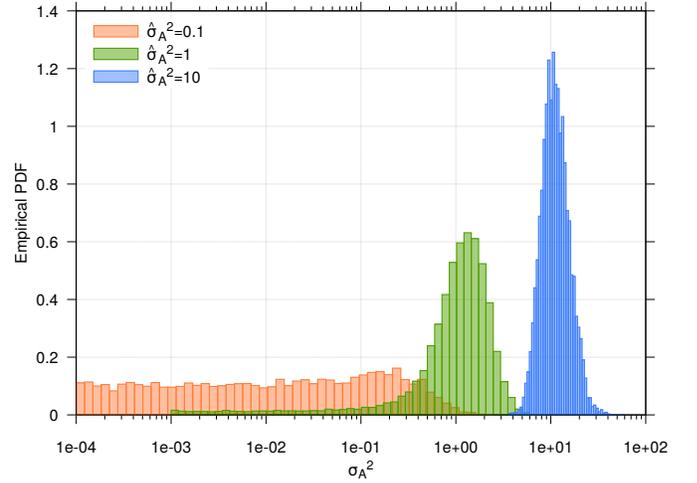}

\vspace{-6mm}

\caption{Normalized histograms of 10\,000 $\sigma_A^2$ parameters giving the final estimate triplets (0.1, 1, 1) in red, (1,1,1) in green, (10,1,1) in blue. The number of EDF is 20.\label{sec:hist_inv_pb}}
\end{figure}

		\subsubsubsection{Case of several EDF}
Before using the simulation results to assess the method accuracy, let us observe them in a few cases: keeping $\fe{B}=\fe{C}=1$, we will successively vary $\fe{A} \in \left\{0.1, 1, 10\right\}$, i.e. corresponding to one final estimate lower, equal \corad{to} or higher than the other two ones.
Figure \ref{sec:hist_inv_pb} shows the histograms of the parameter $\sigma_A^2$ in each of these cases with 20 EDF.

\begin{description}
	\item[$\hat{\sigma}_A^2=0.1$:] \hspace{4mm}  The histogram of the corresponding $\sigma_A^2$ parameter (see Figure \ref{sec:hist_inv_pb} in red) is constant for the lower values, exhibits a very slight bulge between 0.1 and 1, and tends to\engdl{ward} 0 after 1. Obviously, the histogram is limited to $10^{-4}$ at its left-hand side because we limited the random generation to $B_l=10^{-4}$ but this trend should continue down to $\sigma_A^2=0$. The lower bound of the confidence interval should then be 0.
	\item[$\hat{\sigma}_A^2=1$:] \hspace{4mm}  The histogram plotted in green in Figure \ref{sec:hist_inv_pb} shows an important bump between 0.1 and 1 and tends to\engdl{ward} 0 after 1. Nevertheless, there is also a constant tail, although much lower than in the previous case, for the $\sigma_A^2$ values below 0.1 down to 0. Here also this tail is limited by the lower bound of the random generation ($B_l=10^{-3}$). The lower bound of the confidence interval is then still 0.
 	\item[$\hat{\sigma}_A^2=10$:] \hspace{4mm} In this case (see Figure \ref{sec:hist_inv_pb} in blue), $\sigma_A^2$ is well constrained around 10 and the histogram seems to be almost Gaussian. No doubt that the 95 \% confidence interval will be defined for \engdl{both} \engad{the} lower and \engad{the} upper bound.
\end{description}

\begin{figure}[b]
\includegraphics[width=9cm,trim={0 1.8cm 0 1.8cm},clip]{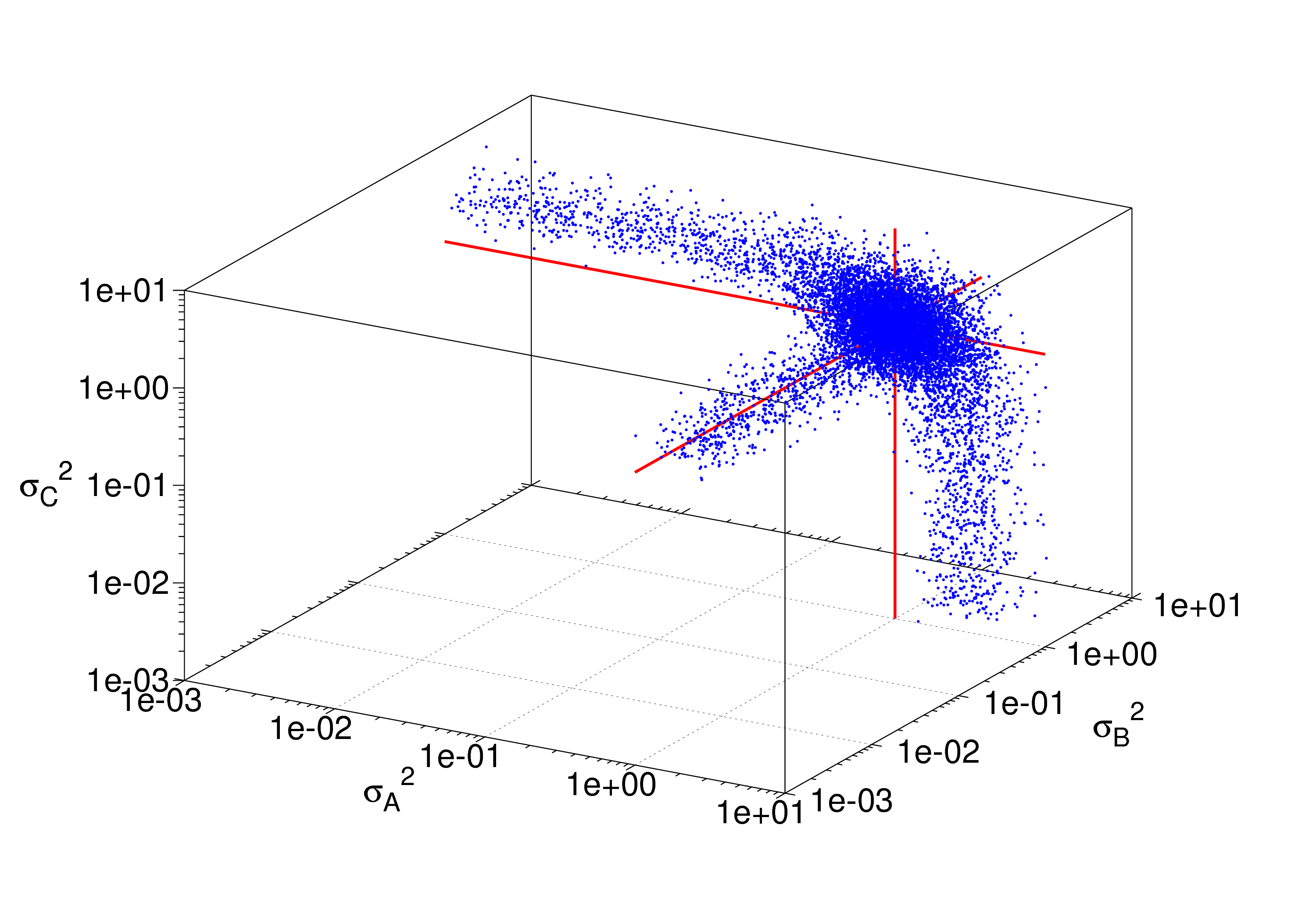}
\caption{3D-plot of 10\,000 parameter triplets giving the final estimate triplets (1,1,1). The number of EDF is 20. The solid red lines represent respectively the lines $\left[\sigma_A^2=\hat{\sigma}_A^2\ \&\ \sigma_B^2=\hat{\sigma}_B^2\right]$, $\left[\sigma_B^2=\hat{\sigma}_B^2\ \&\ \sigma_C^2=\hat{\sigma}_C^2\right]$ and $\left[\sigma_C^2=\hat{\sigma}_C^2\ \&\ \sigma_A^2=\hat{\sigma}_A^2\right]$. \label{sec:3DA}}
\end{figure}

Another representation is given in Figure \ref{sec:3DA}. This 3D plot was built by associating a dimension to each parameter of the triplets from the dataset obtained with a final estimate triplet equal to $(1,1,1)$ (the same dataset as for Figure \ref{sec:hist_inv_pb}). It has the advantage of showing \engdl{well} the relationships between these parameters. This type of plot exhibits a structure with 3 perpendicular branches, more or less dense depending on the final estimate triplet, converging to $(1,1,1)$. 

	\subsubsection{Influence of the number of  EDF\label{sec:acc_EDF}}
In order to check the validity of our method to assess the confidence intervals over the parameters, let us first assign the final estimate triplet to $(1,1,1)$ and vary the number of EDF.
		\subsubsubsection{number \engad{of} EDF varying from 2 to 1000}
\begin{figure}[b]
\centering 
\includegraphics[width=9cm]{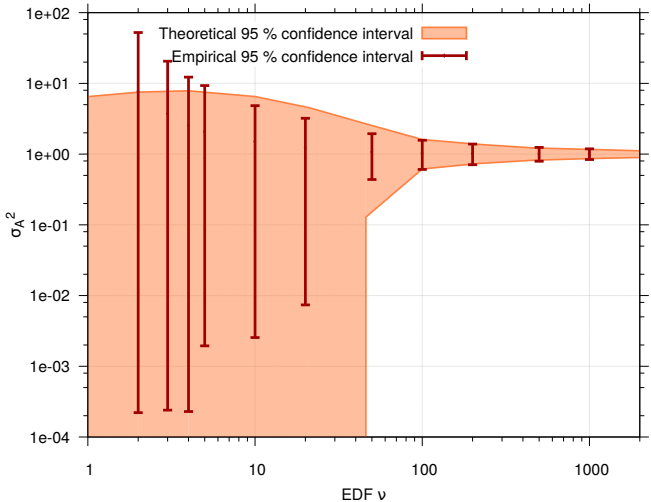}

\vspace{-3mm}

\caption{Influence of the number of EDF $\nu$ on the 95 \% confidence interval over the $\sigma_A^2$ parameter. The final estimate triplet was set at $\hat{\sigma}_A^2=\hat{\sigma}_B^2=\hat{\sigma}_C^2=1$. The colored area corresponds to the assessment of the confidence interval by our method whereas the error bars correspond to the empirical estimation from 10\,000 Monte-Carlo simulations.\label{sec:Avsnu}}
\end{figure}
Figure \ref{sec:Avsnu} shows a pretty good agreement between the 95 \% confidence intervals estimated by our method (colored area) and the one obtained from a set of 10\,000 Monte-Carlo simulations (error bars), except for $\nu=2$ EDF where our method seriously underestimates the 97.5 \% bound. On the other hand, the method is fully reliable from $\nu=5$ EDF and above.

For $\nu$ varying from $2$ to $20$, there is also a noticeable discrepancy between the 2.5 \% bound obtained by our method and by simulations. However, the 2.5 \% bound of the simulation error bars is almost exactly the lower bound $B_l$ which limits the pseudo-random excursion of our simulations. Therefore, this 2.5 \% bound is only due to a computational artifact and the 0 result of our method is more reliable.

For $\nu=50$, our method seems to slightly overestimate the confidence interval\engad{,} but for higher EDF\engdl{,} the Monte-Carlo simulations give\engdl{s} the same confidence intervals \corad{as}\cordl{than} our method. This result is not surprising since the Gaussian approximation we adopted in our algorithm is perfectly justified for large EDFs. 

		\subsubsubsection{Case of 1 EDF}
\begin{table}
\caption{Case of 1 EDF -- Comparison of the 95 \% confidence interval bounds obtained by our method (hereafter LV method standing for Lantz-Vernotte method) and by 10\,000 Monte-Carlo simulations. The final estimate triplet was set at $(-1/2, 1, 1)$.\label{tab:1EDF}}
\begin{center}
\begin{tabular}{c|cc|cc}
\hline
Parameter & \multicolumn{2}{c|}{2.5 \% bound} & \multicolumn{2}{c}{97.5 \% bound}\\
& LV method & Simulation & LV method & Simulation\\
\hline
$\sigma_A^2$ & $0$ & $2.01\cdot 10^{-6}$ & $1.39$ & $159$\\
$\sigma_B^2$ & $0$ & $3.96\cdot 10^{-6}$ & $5.28$ & $474$\\
$\sigma_C^2$ & $0$ & $3.50\cdot 10^{-6}$ & $5.31$ & $464$\\
\hline
\end{tabular}
\end{center}
\end{table}

Table \ref{tab:1EDF} shows that the 97.5 \% bound obtained by our method is \cordl{approximately} underestimated by a factor \corad{of approximately} 100! It should then not be used for $\nu\leq 2$.

	\subsubsection{Influence of relative values of the final estimates}

		\subsubsubsection{Low EDF: $\nu=5$}
\begin{figure}
\includegraphics[width=9cm]{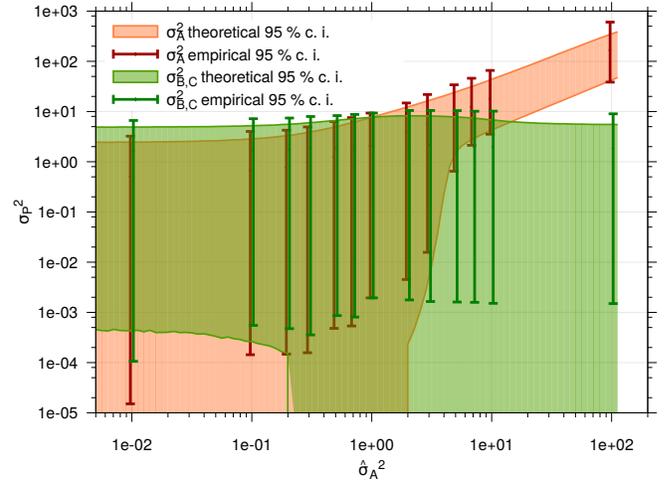}

\vspace{-6mm}

\caption{Influence of the final estimate $\hat{\sigma}_A^2$ on the 95 \% confidence interval over the $\sigma_A^2$ parameter (red) and over the $\sigma_B^2$ or $\sigma_C^2$ parameters (green). The $\hat{\sigma}_B^2$ and $\hat{\sigma}_C^2$ final estimates were set at $1$ and the number of EDF is 5. The colored area correspond to the assessment of the confidence interval by our method whereas the error bars correspond to the empirical estimation from 10\,000 Monte-Carlo simulations. The graphs are plotted using a log-log plot.\label{sec:BCvsA5_log}}
\end{figure}

\begin{figure}
\includegraphics[width=9cm]{nu5_ABC_unc_lin.pdf}

\vspace{-6mm}

\caption{Influence of the final estimate $\hat{\sigma}_A^2$ on the 95 \% confidence interval over the $\sigma_A^2$ parameter (red) and over the $\sigma_B^2$ or $\sigma_C^2$ parameters (green). The $\hat{\sigma}_B^2$ and $\hat{\sigma}_C^2$ final estimates were set at $1$ and the EDF is 5. The colored area corresponds to the assessment of the confidence interval by our method whereas the error bars correspond to the empirical estimation from 10\,000 Monte-Carlo simulations. The graphs are plotted using a linear X-scale.\label{sec:BCvsA5_lin}}
\end{figure}

Figure \ref{sec:BCvsA5_log} shows a very good agreement between our method and the simulations for the parameter $\sigma_A^2$ (in red). There is only one slight underestimation of the 2.5 \% bound for $\hat{\sigma}_A^2=4$. Similarly, for the parameter $\sigma_B^2$ or $\sigma_C^2$ (see Figure \ref{sec:BCvsA5_log} in green), there is a good agreement between our method and the simulations. However, our method shows a 2.5 \% bound which increases for \engad{a} final estimate $\hat{\sigma}_A^2<0.1$ which is, unexpectedly,  not confirmed by the simulations. 

Figure \ref{sec:BCvsA5_lin} shows the same parameters versus the same final estimate $\hat{\sigma}_A^2$ but by using a linear X-scale allowing us to plot the negative values of the final estimate. For $\sigma_A^2$ (red plot on Figure \ref{sec:BCvsA5_lin}), our method gives good results down to $\hat{\sigma}_A^2=-0.3$ but, surprisingly, the 97.5 \% bound increases for $\hat{\sigma}_A^2<-0.3$. Here also, it is not confirmed by the simulations  The same odd behavior may be observed for $\sigma_B^2$ and $\sigma_C^2$ (green plot on Figure \ref{sec:BCvsA5_lin}) for the 2.5 \% bound around $0$ as well as for the 97.5 \% bound below $-0.3$.

\cordl{However, t}\corad{T}hese discrepancies between our method and the simulations \corad{are due to the difference between the true $\chi^2$ distribution with a small number of EDF and the normal law approximation in step S\ref{it:approxLG} of the algorthm (see \S \ref{sec:algo}). However, it} do\corad{es} not really matter since, on one hand, the\cordl{y}\corad{se discrepancies} are limited, and on the other hand, they occur only for final estimates close to the low limit (-0.5, see Section \ref{sec:finest_prop}) and are therefore \corad{rare}\cordl{seldom}.

		\subsubsubsection{High EDF: $\nu=100$}
\begin{figure}
\includegraphics[width=9cm]{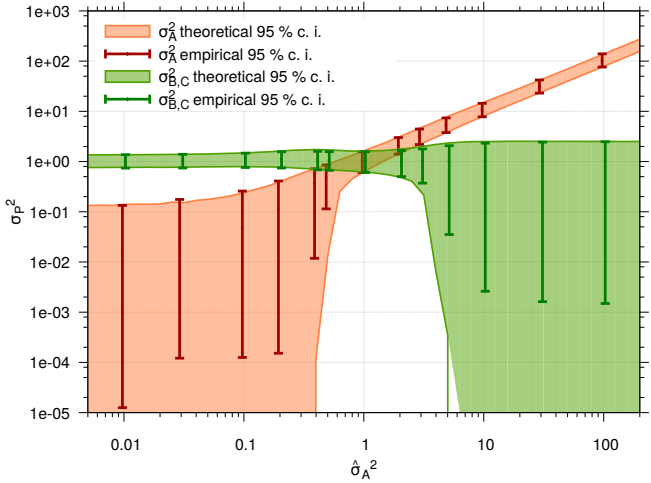}

\vspace{-6mm} 

\caption{Influence of the final estimate $\hat{\sigma}_A^2$ on the 95 \% confidence interval over the $\sigma_A^2$ parameter (red) and over the $\sigma_B^2$ or $\sigma_C^2$ parameters (green). The $\hat{\sigma}_B^2$ and $\hat{\sigma}_C^2$ final estimates was set at $1$ and the EDF is 100. The colored area corresponds to the assessment of the confidence interval by our method whereas the error bars correspond to the empirical estimation from 10\,000 Monte-Carlo simulations. \label{sec:BCvsA20}}
\end{figure}

Figures \ref{sec:BCvsA20} shows that our method works perfectly for 100 EDF. The only noticeable discrepancies concern the 2.5 \% bound and is, once again, clearly due to the computational artifact of the simulation error bars mentioned above.

\section{Conclusion}
\corad{We have performed a thorough theoretical study of the statistics of the 3 cornered hat or Groslambert covariance estimates (direct problem) showing that they 
 are formed by
random variables which are the differences of 2 $\chi^2$ r.v. Massive simulations have been 
 performed 
 and the agreement between these simulations and the results of the theoretical study is very convincing.}

We have \corad{also} proposed a \corad{first attempt} \cordl{method} to assess a confidence interval over the true clock stabilities \corad{(inverse problem) relying upon the 
 determination
 of the posterior probability density function by a Monte-Carlo computation. Here also, we have compared the results of this method to massive simulations}. This \corad{study shows that this} method may be used from EDF larger than 2 but is fully reliable from 5 EDF and beyond. This means that for a data run of duration $T$, 
 the uncertainty domain is valid for an integration time $\tau\leq T/10$ in the case of a white FM noise and below for flicker or random walk FM.

On the other hand, this method is relatively slow since its algorithm relies on a Monte-Carlo scheme involving $10^7$ random draws. This causes a computation time of the order of 1 minute per confidence interval (per error bar).

However, since the analysis of the direct problem presented in this paper seems to be convincingly achieved, it could constitute the basis of a new method of solving the inverse problem, i.e. of estimating confidence intervals around the true stability of the clocks knowing the results obtained by \engad{the} three cornered hat/GCov \engad{method}.

\section*{Acknowledgment}
This work \engdl{is} \engad{was} partially funded by the ANR Programme
d'Investissement d'Avenir (PIA) under the Oscillator IMP
project and the FIRST-TF network. The authors are very grateful to Professor Enrico Rubiola for suggesting us the subject of this work. 

\bibliography{3CH_uncertainties.bib}


\addcontentsline{toc}{section}{Appendix}
\renewcommand\thesubsection{A-\arabic{subsection}}
\renewcommand\thesubsubsection{A-\arabic{subsection}.\arabic{subsubsection}}

\section*{APPENDIX}
	\subsection{Difference of two random variables following two independent $\chi_\nu^2$ distributions\label{sec:annexe}}

		\subsubsection{Definition}
We consider a random variable $X$ which is the difference of two independent random variables \cordl{according to} \corad{following each} a $\chi^2$ law with the same number of degrees of freedom $\nu$:
\begin{equation}
X=\mathcal{A} \dot{\chi}_\nu^2-\mathcal{B} \ddot{\chi}_\nu^2\label{eq:distrib}
\end{equation}
with $\mathcal{A}, \mathcal{B}, \nu \in \mathbb{R}^+$ and $\nu \geq 1$. \corad{The upper dots of $\dot{\chi}_\nu^2$ and $\ddot{\chi}_\nu^2$ make it possible to distinguish 2 independent random variables following the same $\chi^2$ distribution.}

The probabality density function (PDF) $p(x)$ of the random variable $X$ is the variance-gamma distribution:
\begin{equation}
p(x)=\frac{\kappa^{2\lambda} | x - \mu|^{\lambda-1/2} K_{\lambda-1/2} \left(\eta|x - \mu|\right)}{\sqrt{\pi} \Gamma (\lambda)(2 \eta)^{\lambda-1/2}} \; e^{\theta (x - \mu)}\label{eq:pdf}
\end{equation}
with $\eta=(\mathcal{A}+\mathcal{B})/4\mathcal{A}\mathcal{B}$, $\theta=(\mathcal{A}-\mathcal{B})/4\mathcal{A}\mathcal{B}$, $\lambda=\nu/2$, $\mu=0$, $\kappa=\sqrt{\eta^2-\theta^2}$ and $K_\omega(z)$ is a hyperbolic Bessel function of second kind ($\omega \in \mathbb{R}$ and $z\in\mathbb{C}$)\footnote{References : \url{https://math.stackexchange.com/questions/85249/distribution-of-difference-of-chi-squared-variables} and \url{https://en.wikipedia.org/wiki/Variance-gamma_distribution}. The relationships between the $\eta, \theta$ coefficients and the $\mathcal{A}, \mathcal{B}$ parameters have been empirically determined.}.




	\subsubsection{Simulation}
The random variable $X$ following the distribution defined in (\ref{eq:distrib}) was simulated with $\mathcal{A}=1$, $\mathcal{B}=1/3$ and $\nu=5$ :
$$
X=\dot{\chi}_5^2-\frac{1}{3}\ddot{\chi}_5^2.
$$
A number of $N=10\,000\,000$ random draws was realized.

The probability density expressed in (\ref{eq:pdf}) was compared to the histogram obtained from $N$ draws. 
The agreement is almost perfect. 

Finally, the CDF fractiles were calculated for 2.5 \% and 97.5 \% to achieve a 95 \% confidence interval. 
Here too, the agreement is excellent.

A similar study was conducted for the $(\mathcal{A}, \mathcal{B}, \nu)$ triplets in $\left\{(1,1/3,5),(1,1/2,1),(1,1/4,1),(2,2,1),(1,1,1)\right\}$. The results were equally concordant.

	\subsection{Calculation of the true covariance matrix of the estimates}
We suppose in this appendix that the true variances are known. With this hypothesis of the model world, the estimates, either final or elementary, become random variables, with means equal either to the true variances (final estimates) or to the sum of two variances (elementary estimates). We calculate in this appendix their covariance matrix, that is used to solve the inverse problem.

		\subsubsection{Covariance matrix of the elementary estimates\label{sec:covmat}}
From now on, we will denote $\xp(\cdot)$, $\var(\cdot)$ and $\cov(\cdot)$ respectively the mathematical expectation, the variance and the covariance of the quantity $\cdot$ which stands here for $\sz$ measurements as well as elementary or final estimates (variance of variances!).

To be specific, we calculate in the following a diagonal element, $\mathrm{Var}(\hat{\sigma}_{AB}^2)$, and a non diagonal element, $\mathrm{Cov}(\hat{\sigma}_{AB}^2,\hat{\sigma}_{AC}^2)$. We assume that successive measurements of the same quantity are independent, meaning that the variances and covariances after $m$ measurements are equal to the (co)variances after one measurement divided by $m$. Hence, the following calculations are presented with $m=1$.  For a diagonal element, we obtain:
\begin{eqnarray}
\mathrm{Var}(\hat{\sigma}_{AB}^2)&=&\xp\left[(\sz_A-\sz_B)^4\right] - \left\{\xp\left[(\sz_A-\sz_B)^2\right]\right\}^2 \nonumber\\
&=&\mathrm{Var}(\sz_A^2)+\mathrm{Var}(\sz_B^2)+4\mathrm{Var}(\sz_A)\cdot\mathrm{Var}(\sz_B)\nonumber
\end{eqnarray}
We assume that $\sz_A$ is centered Gaussian, meaning that $\sz_A^2$ follows a $ \chi^2$ law of mean $\mathrm{Var}(\sz_A)$. For such a law, we have $\mathrm{Var}(\sz_A^2)=2\left[\mathrm{Var}(\sz_A)\right]^2$, which allows a more compact formulation of the above result:
\begin{eqnarray}
\mathrm{Var}(\hat{\sigma}_{AB}^2)&=& 2 \left[\mathrm{Var}(\sz_A)\right]^2+ 2 \left[\mathrm{Var}(\sz_B)\right]^2\nonumber\\
&&+4\mathrm{Var}(\sz_A)\cdot\mathrm{Var}(\sz_B)\nonumber\\
&=&2 \left[\mathrm{Var}(\sz_A)+\mathrm{Var}(\sz_B)\right]^2.\nonumber
\end{eqnarray}

We pass now to a non diagonal element:
\begin{eqnarray}
\mathrm{Cov}(\hat{\sigma}_{AB}^2,\hat{\sigma}_{AC}^2)&=&\xp\left[(\sz_A-\sz_B)^2(\sz_A-\sz_C)^2\right] \nonumber\\
&&- \xp\left[(\sz_A-\sz_B)^2\right]\cdot\xp\left[(\sz_A-\sz_C)^2\right] \nonumber\\
&=&2\left[\mathrm{Var}(\sz_A)\right]^2.\nonumber
\end{eqnarray}

\subsubsection{Covariance matrix of the final estimates}
We use the Groslambert covariance definition of the final estimates, 
 $\hat{\sigma}_{P}^2=\widehat{\mathrm{GCov}}_{PO,PQ} $ and calculate, as above, the covariance of these estimates: $\mathrm{Cov}(\hat{\sigma}_{P}^2,\hat{\sigma}_{Q}^2)$. We assume no measurement noise. As in the previous subsection, we consider the case $m=1$ measurement and we give explicit clock symbols to $P$ and $Q$: first, we calculate the diagonal elements of the estimates covariance matrix by choosing $P=Q=A$, then the non diagonal elements by choosing $P=A,\ Q=B$.  In the first case, we obtain:
\begin{eqnarray}
\mathrm{Var}(\hat{\sigma}_{A}^2)&=&\xp\left\{\left[(\sz_A-\sz_B)(\sz_A-\sz_C)\right]^2\right\} \nonumber\\
&&- \left\{\xp\left[(\sz_A-\sz_B)(\sz_A-\sz_C)\right]\right\}^2 \nonumber\\
&=& \mathrm{Var}(\sz_A^2)+\mathrm{Var}(\sz_B)\cdot\mathrm{Var}(\sz_C)\nonumber\\
&&+\mathrm{Var}(\sz_A)\cdot\mathrm{Var}(\sz_B)+\mathrm{Var}(\sz_A)\cdot\mathrm{Var}(\sz_C)\nonumber
\end{eqnarray}
The non diagonal case gives:
\begin{eqnarray}
\mathrm{Cov}(\hat{\sigma}_{A}^2,\hat{\sigma}_{B}^2)&=&\xp\left[(\sz_A-\sz_B)(\sz_A-\sz_C)\right.\nonumber\\
&&\left.\cdot(\sz_B-\sz_A)(\sz_B-\sz_C)\right] \nonumber\\
&&- \xp\left[(\sz_A-\sz_B)(\sz_A-\sz_C)\right]\nonumber\\
&&\cdot\xp\left[(\sz_B-\sz_A)(\sz_B-\sz_C)\right] \nonumber\\
&=&\mathrm{Var}(\sz_A)\cdot\mathrm{Var}(\sz_B)\nonumber\\
&&-\mathrm{Var}(\sz_C)\left[\mathrm{Var}(\sz_A)+\mathrm{Var}(\sz_B)\right]\nonumber
\end{eqnarray}

In both cases, we have used the independence of the oscillators, giving: 
$\xp\left(\sz_P^2\sz_Q^2\right)=\xp\left(\sz_P^2\right)\cdot\xp\left(\sz_Q^2\right)$ and their zero mean, giving $\mathrm{Var}(\sz_P)=\xp\left(\sz_P^2\right)$.

Using the elementary or the final estimates gives exactly the same results, for the same measurements and the same Monte-Carlo set.

\end{document}